\documentstyle[prl,aps,preprint]{revtex}
\input{epsf}

\def\p{{\bf p}}
\def\k{{\bf k}}

\def\r{{\bf r}}
\def\rp{{\bf r^\prime}}

\def\G{{\bf G}_n}
\begin{document}

\title{ Treatment of Correlation Effects in Electron Momentum Density:
Density Fuctional Theory and Beyond}

\author{
B. Barbiellini and 
A. Bansil
}

\address{Department of Physics, Northeastern University, Boston, 
MA 02115 USA}

%\date{\today}
\maketitle
\begin{abstract}

Recent high resolution Compton scattering experiments clearly reveal that 
there are fundamental limitations to the conventional local density 
approximation (LDA) based description of the ground state electron 
momentum density (EMD) in solids. In order to go beyond the framework
of the density functional theory (DFT), we consider for the correlated 
system a BCS-like approach in which we start with a singlet pair 
wavefunction or a 'geminal' from which the many body wavefunction is 
then constructed by taking an antisymmetrized geminal product (AGP). 
A relatively simple practical implementation of the AGP method
is developed where the one-particle orbitals are approximated by the 
Kohn-Sham solutions used in standard band computations, and the 
orbital-dependent BCS energy scale $\Delta_i$ is determined through a 
readily computed exchange-type integral. The methodology is illustrated 
by considering EMD and Compton profiles in Li, Be and Al. It is found that 
in Li the present scheme predicts a substantial renormalization of the 
LDA result for the EMD; in Be, the computed correlation effect is 
{\em anisotropic}, while in Al, the 
deviations from the LDA are relatively small. These theoretical 
predictions are in qualitative accord with the corresponding 
experimental observations on Li, Be and Al, and indicate the 
potential of the AGP method for describing correlation effects on the 
EMD in wide classes of materials.

\end{abstract}


\newpage
\section{Introduction}

Recent high resolution Compton scattering experiments on a number 
of materials clearly show that the EMD of the ground state is not
described satisfactorily by the conventional LDA-based framework
\cite{li1,li2,li3,li4,be1,be2,be3,al1,al2,alli,cu}. 
In Li, the size of the discontinuity $Z_k$ in the EMD at the Fermi 
momentum $p_F$ appears to be anomalously small  
\footnote {We will not concern ourselves here with 
possible complications
in the interpretation of Compton data with regard to questions of 
background subtraction, failure of impulse approximation, etc.}
 --nearly zero \cite{li1,li2,li3,li4}, 
and differs sharply from the values of $0.55-0.85$ deduced 
via a variety 
of studies of the correlated homogeneous electron gas (HEG) stretching 
over last several decades
\cite{daniel,lundqvist,over71,lantto80,farid,qmc_heg}. 
In Be, comparisons between highly accurate 
computed and measured Compton spectra 
\cite{be3}
indicate systematic discrepancies 
which are {\em anisotropic} and would be difficult to explain within
the standard Lam-Platzman (LP) type correlation correction to the 
EMD \cite{phil,bauer,cardwell}
which is {\em isotropic} by construction. The EMD in Al, on the 
other hand, is adduced to be reasonably close to the LDA predictions
\cite{al1,al2,alli}. 
Several attempts have been made to gain a theoretical handle on this 
problem \cite{kubo96,shulke99,eguiluz00,louie,filippi}. 
The early optimism of a GW computation \cite{kubo96} in Li did not hold up to 
a later investigation \cite{shulke99,eguiluz00}. 
Surprisingly, Quantum Monte-Carlo (QMC) 
studies of Si and Li \cite{louie,filippi} also do not reveal any 
substantial differences in 
the EMD with respect to the LDA predictions. It should be emphasized that a 
successful theory must not only explain what may be thought of as 
the "low" density case of monovalent Li, but also
it must approach the LDA results 
correctly in the "high" density limit of trivalent Al in addition
to explain the behavior of divalent Be.

In order to make progress, we observe first that the current GW and 
QMC computations in connection with the EMD assume that the fermionic 
correlation is not modified drastically by interactions and that 
an adiabtic path exists between the free and the interacting electron 
gas. The many body wavefunction underlying the GW and QMC work
is built from Slater determinants of single-particle orbitals, 
with an implicit nodal structure and plasmon-type physics of 
correlations \cite{fulde} which is more or less similar to that of the LDA
\cite{review_dft}. 
Bearing these considerations in mind, we have been motivated 
to examine other strategies, and here we propose a
BCS-like approach for going beyond the framework of the DFT 
\cite{blatt1,blatt,bratoz,bouchard,agp0,agp1},
wherein we start with singlet electron 
pairs-- referred to as `geminals' \cite{coleman}, and 
then combine these objects into 
an antisymmetrized product (AGP) to obtain the wavefunction of the 
many body system. Such an idea has been invoked previously, for example, 
in connection with liquid He$^3$ \cite{bouchard} and in quantum chemistry 
calculations of molecules \cite{bratoz}. An application to the problem of the 
EMD in extended systems which is presented here has, to our 
knowledge, not been reported in the literature before. Notably, 
the AGP wavefunction in molecules typically yields only $40-50$ \% 
of the available correlation energy \cite{bratoz}. However, 
our main goal here is to develop a working scheme which can capture 
some of the essential physics of the EMD of the 
correlated electron gas, and not so much to obtain the total
energy accurately. In this spirit, it is hoped that the AGP method 
can provide a useful tool for understanding the recent Compton spectra
and the nature of the EMD over a wide range of electron densities. 

An outline of this article is as follows. We start in Section II with a 
rigorous expression for the EMD, $\rho(\bf {p})$, in the correlated 
electron gas in terms of the eigenvalues and eigenfunctions of the 
one-particle density matrix $\hat{\rho}(\r,\r')$ \cite{lowdin,goscinki}.
Section III discusses the 
conventional Lam-Platzman (LP) treatment \cite{phil,bauer,cardwell}
of correlations within the LDA
which is based on the momentum density $\rho_h(p)$ of the interacting 
HEG \cite{daniel,lundqvist,over71,lantto80,farid,qmc_heg}, 
including a useful parametrization of the 
more recent QMC data for $\rho_h(p)$ \cite{qmc_heg}.
Section IV takes up a description 
of the AGP wavefunction and the associated total energy functional
\cite{agp0}. 
A relatively simple practical implementation of the AGP scheme is 
discussed in Section V. The theory is illustrated by considering aspects
of the EMD and Compton profiles in Li, Be and Al in Section VI. 
We emphasize that although these results indicate that the present 
method is promising, further work is necessary for confronting the 
AGP model quantitatively with experiments. Finally, Section VII 
summarizes the main results of this article.

\section{Momentum Density and Natural Orbitals}
\label{sec:theo}

We start with the one-particle density matrix, $\hat{\rho}(\r,\rp)$,
which is defined in terms of
the normalized N-particle wavefunction, $\Psi$, as
\begin{equation}
\hat{\rho}(\r,\rp)=N\int d\xi ~\Psi^*(\r,\xi)\Psi(\rp,\xi)~, 
\nonumber
\end{equation}
where the integral extends over the coordinates of all other particles. 
The eigenfunctions and eigenvalues of 
$\hat{\rho}$ define the {\em natural orbitals}, $\psi_i$, and the 
associated occupation numbers, $n_i$ \cite{lowdin}. 
The density operator $\hat{\rho}$ can then be expanded into projectors 
$ |~\psi_i><~\psi_i~|$ via the spectral theorem:
\begin{equation}
\hat{\rho}=\sum_{i=1}^{\infty} n_i~|~\psi_i><~\psi_i~|~.
\end{equation}
If one requires the natural orbitals to possess the symmetry of the
Hamiltonian, then they constitute a unique decomposition 
in an orthonormal basis set. 
Other basis sets for describing the many body wavefunction are often used, 
e.g. the {\em generalized overlap amplitudes} \cite{goscinki} which 
are linearly dependent. Any basis set can of course be related 
to the natural orbitals through an appropriate canonical transformation.

The EMD is defined as the momentum transform of the density matrix, 
\begin{equation}
\rho(\p)= \frac{1}{8 \pi^3}
\int \int d^3\r d^3\r'
\hat{\rho}(\r,\rp) \exp(-i \p \cdot (\r - \rp))~.
\end{equation}
Note that, in general, the EMD involves off-diagonal elements of the 
real space density matrix. In terms of the natural orbitals $\psi_i$,
however, the EMD can be cast in the simple form
\begin{equation}
\rho(\p)=\sum_i n_i~|<\p~|~\psi_i>|^2~, 
\end{equation}
where $<\p~|~\psi_i>$ is the momentum transform of $\psi_i$.

If the many body wavefunction is represented by a single 
determinant, which is true in the case of Hartree Fock or the DFT
\cite{kohn96}, 
then the density matrix is idempotent $(\hat{\rho}=\hat{\rho}^2)$ 
and reduces to a summation over the occupied spin-dependent 
orbitals $\psi_{i}$, i.e.
\begin{equation}
\hat{\rho}({\bf r},{\bf r}^\prime) = 
 \sum_{i=1}^{N} \psi_i({\bf r})\psi^*_i({\bf r}^\prime)\,.
\end{equation}
The EMD in this {\em independent particle model} (IPM) is 
similar to the more general Eq. (2), except that the occupation numbers 
are now strictly 0 or 1 depending upon whether 
the state in question is empty or filled. 
Electron correlations, in effect, allow these occupied and empty IPM 
states to mix so that the occupation of states below the Fermi 
energy ($E_F$) becomes less than 1, while that of states above $E_F$ 
becomes non-zero.

One further point is noteworthy. In a periodic system, the electronic 
states are the Bloch states, $\psi_{\k \mu}(\r)$, where 
$\mu$ is the band index and $\k$ is the crystal momentum 
which is restricted to the first Brillouin zone (BZ).
The associated occupation numbers, $n_i=N_{\mu}(\k)$, 
possess translational symmetry, i.e. 
\begin{equation}
N_{\mu}(\k)=N_{\mu}(\k+\G), 
\label{eq:trans}
\end{equation}
with respect to the set of reciprocal lattice vectors $\G$.

\section{Correlation Correction in the Density Functional Theory}

As a consequence of the Hohenberg-Kohn theorem \cite{review_dft}, 
the ground-state expectation value of any operator $\hat{O}$ 
is a functional $O[n(\r)]$ of the electron
density $n(\r)$. It can be shown that \cite{bauer}
\begin{equation}
O[n(\r)]=O_0[n(\r)] + 
\frac{d}{d\lambda} E_{xc}[n(\r)](\lambda)~,
\label{eq:bauer}
\end{equation}
where $O_0[n(\r)]$ is the expectation value 
for a system of noninteracting fermions moving
in the effective field involved in the Kohn-Sham formalism,
$E_{xc}$ is the exchange-correlation energy functional and
$\lambda$ is a scalar coupling parameter for the operator $\hat{O}$.
The value obtained for the non-interacting 
case-- given by the first term on the right side of 
Eq. (\ref{eq:bauer}), must be 
corrected by the second term. This so-called LP 
correction\cite{phil} provides a formal scheme for treating 
correlation effects in the interacting system within the DFT. 

Insofar as the momentum density in extended systems is 
concerned, LDA is perhaps the most common implementation 
of the DFT. The relevant expression for the 
LP correction, $\Delta\rho[n(\r)](p)$, is based on the 
treatment of $E_{xc}$ in the HEG \cite{bauer}. 
\begin{equation}
\Delta \rho [n(\r)](p)=\int_{\mbox{WS}}~
               [\rho_h(r_s({\bf r}),p)-\rho_0(p)]
               ~n({\bf r})~d^3{\bf r}~.
\label{EqLP}
\end{equation}
Here, the integral extends over the Wigner-Seitz (WS) unit cell. 
The square brackets in the integrand give the difference between the momentum
densities $\rho_h$ and $\rho_0$ of the interacting and non-interacting
HEG, respectively, evaluated at the local density $n(\r)$ of the physical
system given by the electron density parameter $r_s(\r)$. 

Despite the limitations of form (\ref{EqLP})  --a 
point to which we return below, this equation has been used 
extensively in the literature for evaluating the LP correction 
by using a variety of different 
results for the HEG starting with the work of Daniel and Vosko
\cite{daniel,lundqvist,over71,lantto80,farid,qmc_heg}. 
The most recent QMC data for $\rho_h(p)$ however 
would be the most satisfactory to employ \cite{qmc_heg}.

\begin{figure}[htb]
\unitlength=1cm
\begin{center}
\begin{picture}(7,8.3)
\put(-2.0,-1.0){\epsfysize=9cm
\epsffile{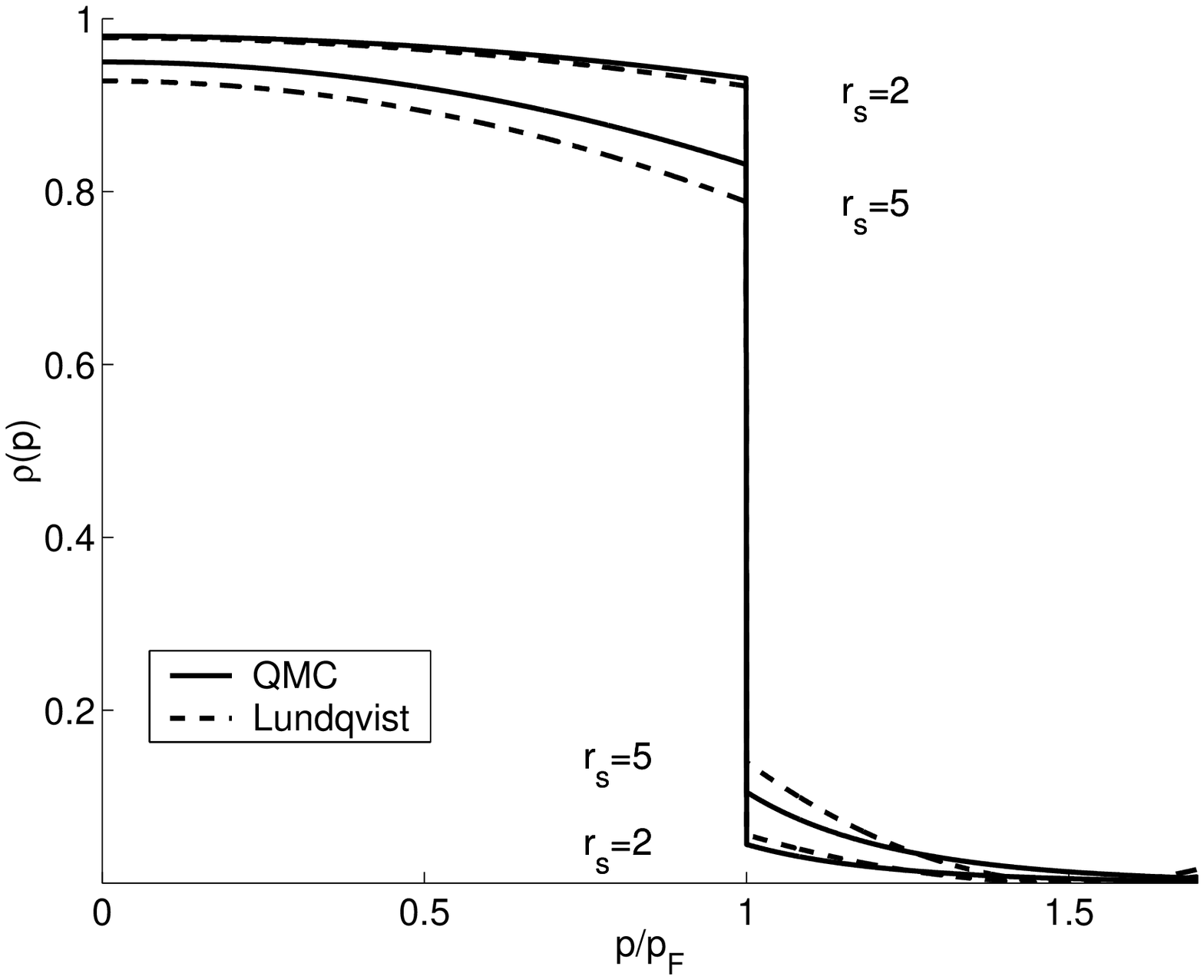}}
\end{picture}
\end{center}
\vskip 0.5cm
\caption{\ }
%\caption{
%Momentum density $\rho(p)$ in the homogeneous electron gas. 
%The momentum $p$ is given in scaled units of the free electron Fermi
%momentum $p_F$. Parameterized QMC (solid) and LDA (dashed) results 
%are shown for $r_s$
%values of 2 and 5, as discussed in the text. 
%}
\label{HEG}
\end{figure}

With this motivation, we provide a convenient formula for 
$\rho_h(r_s,p)$ as a function of $r_s$, which we have 
obtained by parameterization of the QMC data
\cite{qmc_heg}. For HEG of density $n$
\vskip 0.3 cm
\begin{equation}
r_s = ({3\over 4\pi n})^{1/3}~.
\end{equation} 
\vskip 0.3 cm
We write QMC results for $\rho_h(p)$ as
\begin{equation}
\label{eq19}
\rho_h(r_s,p)=\cases{a(1)~(1-a(2)~x^2), &if $x\le 1$; \cr
                a(3)~\exp(-a(4)~(x-1))+ T/x^8 &otherwise, \cr}
\label{paraqmc}
\end{equation} 
where $x=p/p_F$ with $p_F=(9/4\pi)^{1/3}/r_s$, and 
a(1) through a(4) are $r_s$-dependent fitting parameters. 
Our form of $\rho_h(r_s,p)$ differs somewhat from that given 
by Farid et {\em al.} \cite{farid}, in that the tail in 
Eq. (\ref{paraqmc}) is exponential while Ref. \cite{farid} uses
a Gaussian tail. The coefficient $T$ is given by Yasuhara and Kawazoe
\cite{yasuhara} as 
\begin{equation}
T=\frac{4}{9}~(\frac{\alpha~r_s}{\pi})^2~g(0)~,
\end{equation}
where $\alpha=(4/9\pi)^{1/3}$ 
and 
\begin{equation}
g(0)=\frac{64}{(8+5~r_s+383/1200~r_s^2)^2}~,
\end{equation}
is the pair correlation function at $r=0$, 
which has been obtained analytically by 
Overhauser \cite{overhauser}.
Finally, we have fitted the coefficients a(1) through a(4)
to the QMC data and obtained the following values. 
\begin{eqnarray}
a(1)&=&1-.010~r_s, \nonumber\\
a(2)&=&0.025~r_s, \nonumber\\
a(3)&=&32/13~\delta, \nonumber\\
a(4)&=&4, 
\end{eqnarray}
where $\delta$ 
\begin{equation}
\delta=1/3-a(1)~(1/3-1/5~a(2))-\frac{T}{5}
\end{equation}
has been chosen to satisfy the electron number sum rule. Fig. \ref{HEG}
shows that the QMC results parametrized here 
over the typical metallic range of densities ($r_s$ varying 
between 2 and 5) differ significantly from those given by 
the early work of Lundqvist \cite{lundqvist} parametrized
by Cardwell and Cooper \cite{cardwell}; the original QMC data 
points lie indistinguishably close to the fitted solid curve 
of Fig. 1 and are not shown for simplicity. 

%\begin{figure}[htb]
%\unitlength=1cm
%\begin{center}
%\begin{picture}(7,8.3)
%\put(-2.0,-1.0){\epsfysize=9cm
%\epsffile{./FIG0.eps}}
%\end{picture}
%\end{center}
%\vskip 0.5cm
%\caption{(a): Parametrization of the QMC results 
%in the HEG for $r_s$
%values of 2,3,4 and 5, as discussed in the text. 
%$r_s$=2 corresponds to the highest curve below the Fermi momentum $p_F$
%and the lowest curve above $p_F$; (b) is the same as (a), except that it 
%refers to the LDA. }
%\label{HEG}
%\end{figure}

\section{Treatment of Correlation Effects Beyond the DFT}

The LP expression of Eq. (\ref{EqLP}) suffers from a number of limitations. 
Most importantly, this form only describes an {\em isotropic} 
redistribution of the EMD, even though the effect of correlations 
will in general be {\em anisotropic}. This also implies that the 
LP corrected EMD will possess a spurious feature at the free 
electron Fermi radius in all directions in the momentum space. 
Finally, the movement of spectral weight from below to above 
$E_F$ in Eq. (\ref{EqLP}) is not very transparent in the sense that it is 
not associated directly with changes in the occupation numbers 
of various states, making it very difficult  
to impose reasonable physical conditions on the occupation numbers
such as those of translational symmetry of Eq. (\ref{eq:trans}).

Bearing these considerations in mind, and the fact that the QMC and 
GW results based on the LDA in Li \cite{eguiluz00,filippi} are not all that 
different from the LP correction and, therefore, would not explain 
the experimental EMD in Li, it is clear that there is need to go 
beyond the basic form of the Slater-Jastrow many-body wavefunction 
implicit in much of the QMC work given by
\begin{equation}
\Psi=F D^{\uparrow} D^{\downarrow},
\end{equation}
where the Jastrow factor $F$ accounts for the plasmon zero 
modes\cite{fulde} and $D^{\uparrow}$ and $D^{\downarrow}$ are
Slater determinants for up and down spin electrons. To make a 
headway, we propose using a BCS-like many body wavefunction 
in which individual elements involve {\em singlet electron pairs}, 
rather than one-particle orbitals. The underlying nodal structure 
of such a many body wavefunction tends to minimize the effects of 
the exclusion principle \cite{blatt1}. 
As already noted, such an approach has been invoked in 
a variety of problems in the literature over the years
\cite{bratoz,bouchard,agp0,agp1}, and here 
we are suggesting its use in connection with the treatment of 
momentum density in extended systems.

The starting point is a singlet pair wavefunction or the associated 
{\em generating geminal} $\phi(\r_1\uparrow,\r_2\downarrow)$ 
\cite{agp0,coleman},
which can be expressed in a diagonal expansion of natural spin orbitals
with undetermined coefficients $g_i$'s:
\begin{equation}
\phi(\r_1\uparrow,\r_2\downarrow)=
\sum_i~g_i~\psi_i^*(\r_1) \psi_i(\r_2)~
[|\uparrow>|\downarrow>-|\downarrow>|\uparrow>]. 
\end{equation}
The many body wavefunction then is the AGP
\begin{equation}
\Psi=\hat{A} \left \{ \Pi_{i=1}^{N/2} 
\phi(\r_{i} \uparrow,\r_{j} \downarrow) \right \}~,
\label{eq:agp1}
\end{equation}
where $\hat{A}$ is the antisymmetrization operator.
The AGP belongs to the category of wavefunctions of {\em extreme type}
\cite{coleman}. \footnote{
Incidentally, if $N$ is odd,
the extreme type wavefunction is proportional to the antisymmetrized
product of a one body orbital $f$ and $(N-1)/2$ factors of $\phi$; $f$
must be strongly orthogonal to $\phi$, that is, 
$\int f(\r_1) \phi(\r_1,\r_2) d^3\r_1=0$ for all $\r_2$ \cite{coleman}.}
Eq. (\ref{eq:agp1}) can be cast into an $N/2 \times N/2$
determinant
\begin{equation}
\Psi=\mbox{Det}|\phi(\r_{i} \uparrow,\r_{j} \downarrow)|.
\label{eq:agp2}
\end{equation}
The one-body wavefunctions $\psi_i$ are the natural orbitals
for not only the generating geminal $\phi$, but also for the 
AGP wavefunction $\Psi$ \cite{agp0}. 
For a two electron system, AGP is exact and
equivalent to a configuration interaction calculation.

The $\psi_i$'s and $g_i$'s may be determined by minimizing the total 
energy functional. In extended systems the coefficients $g_i$ are 
not particularly convenient, and it is more useful to introduce 
a new set of coefficients $h_i$ to define a "Cooper pair" function given by:
\begin{equation}
C(\r_1\uparrow,\r_2\downarrow)=
\sum_i~h_i~\psi_i^*(\r_1) \psi_i(\r_2)~
[|\uparrow>|\downarrow>-|\downarrow>|\uparrow>]
\end{equation}
The non-Hartree-Fock-like term in the total energy can then be 
written as \cite{agp0} 
 \begin{equation}
E[h_i,\psi_i]=E_{HF}[\hat{\rho}] +  E_{BCS}[h_i,\psi_i] + O(1/N)~,
\label{toten}
\end{equation}
where $E_{HF}$ is the Hartree Fock functional and
$E_{BCS}$ is a BCS-type functional \cite{blatt,agp0}
given by
\begin{equation}
E_{BCS}=\frac{1}{2} <C|V_{12}|C>~.
\end{equation}
The normalization of the AGP wavefunction 
imposes a relationship between the coefficients 
$h_i$ and the occupation numbers $n_i$ via the condition
\cite{agp0}
\begin{equation}
h_i=\pm\sqrt{n_i(1-n_i)}.
\label{eq:h}
\end{equation}
Note that for Coulomb interaction the pair potential $V_{12}$ is 
repulsive, so that energy can be gained only through the 
exchange part $E_{HF}$ of the Hartree-Fock functional. It can be shown that 
the amplitudes $h_i$'s then must change sign at the "pseudo-Fermi surface"
corresponding to the Hartree-Fock solution (i.e. the solution that 
neglects the term $E_{BCS}$) \cite{agp0}.
The energy can also be gained of course through the term $E_{BCS}$ 
in Eq. (\ref{toten}) by the introduction of lattice dynamics as is the case at 
the superconducting transition \cite{weger}.

Recently, Goedecker and Umrigar \cite{gu1,gu2} have considered
an approximation to the two-particle density matrix $\sigma$
which leads to a functional of natural orbitals similar to form 
(\ref{toten}) discussed here. However Cs\'anyi and Arias \cite{gabor} 
have shown that the electronic states predicted by Ref. \cite{gu1}
are overcorrelated; the reason is that $\sigma$ is varied over too large
a class of functions without the restriction of $N$-representability. 
This problem is circumvented in our case since the 
the AGP functional is $N$-representable by construction.

\section{A Simple Implementation of the AGP Model}
\label{sec:calc}

Our main goal is to gain a handle on the nature of occupation numbers 
in the correlated electron gas. In this spirit, we start by approximating 
the natural orbitals $\psi_i$ by the Kohn-Sham orbitals \cite{ks} for the 
one-particle crystal potential; the associated eigenvalues are 
denoted by $\varepsilon_i$ with $\varepsilon_i=0$ defining the 
Fermi level following the usual convention. 

The solution to the BCS problem involves an energy scale $\Delta_i$
\cite{blatt} which determines the mixing of states above
and below the Fermi level. The occupation numbers 
are given by
\begin{equation}
n_i=\frac{1}{2} \left (
1 - \frac{\varepsilon_i}
{\sqrt{\varepsilon_i^2+\Delta_i^2}}
\right )~.
\label{renocc}
\end{equation}
$n_i$ is obviously greater than $1/2$ for $\epsilon_i < 0$, and less than
$1/2$ for $\epsilon_i > 0$; for $\epsilon<<\Delta_i$, 
$n_i\longrightarrow 1$ and for 
$\epsilon>>\Delta_i$, $n_i\longrightarrow 0$. At 
$\varepsilon_i=0$, $n_i=1/2$. 
In order to make progress at this point we require a method for estimating
the value of $\Delta_i$. As already noted, there will be cancellation between
the terms $E_{BCS}$ and $E_{HF}$ in Eq. (\ref{toten}), and we would 
physically expect
the total energy of the AGP wavefunction to be quite close to that of the 
Kohn-Sham Hamiltonian. With this motivation, we make the assumption that
{\em the energy cost of electron pairing will, in fact, be roughly compensated
by the exchange energy}. If so, it readily follows that
\begin{equation} 
|\Delta_i| \sim I_i~,
\label{eq:exch}
\end{equation} 
where 
\begin{equation}
I_i=\left |
\int d^3\r_1 d^3\r_2
\frac{\delta^2 E_x}{\delta n(\r_1) \delta n(\r_2)}~
|\psi_i^*(\r_1) \psi_i(\r_2)|^2
\right | 
\end{equation}
is an exchange type integral \cite{janak}.
We emphasize that the validity of 
assumptions leading to Eq. (\ref{eq:exch}) is unclear and that 
this point deserves 
further investigation. Within the LDA, an explicit formula for 
$I_i$ can be obtained straightforwardly: 
\begin{equation}
I_i=\frac{1}{3} \int d^3\r 
|\psi_i(\r)|^4 \frac{v_x( \r)}{n(\r)}~,
\label{eq:janak}
\end{equation}
where $v_x( \r)=2/\pi(3\pi^2n(\r)^{1/3}$ is the Kohn-Sham
exchange potential \cite{ks}. As expected, $I_i$ is large when 
$\psi_i(\r)$ is confined to a small spatial region. 
Finally, the $n_i$'s are renormalized in order to satisfy 
the correct sum rule, i.e. 
\begin{equation}
N=\sum_i n_i~.
\end{equation}
The evaluation of the EMD follows along the lines of the standard IPM 
computations, except that the occupation numbers for the correlated
electron gas given by Eq. (\ref{renocc}) are used. For the specific EMD and 
Compton profile results on Li, Be and Al reported in Section VI below, 
we have used orbitals $\psi_i$ calculated using linear muffin-tin 
orbital (LMTO) \cite{lmto1,lmto2} band structure method for convenience, 
including corrections to the EMD due to the overlapping sphere 
geometry\cite{lmto_emd}.

The essential parameter which controls the correlation effect in 
our theory is,
\begin{equation} 
\alpha= \Delta/w \sim I_i/w,
\label{eq:alpha}
\end{equation}
where $w$ denotes the relevant valence electron 
band-width. By keeping Eqs. (\ref{renocc}) and 
(\ref{eq:exch}) in mind, it is evident that when $I_i$ is comparable to $w$, 
$\alpha \sim 1$, 
and states deep in the Fermi sea are renormalized inducing significant 
shift of spectral weight from below to above $\varepsilon_i=0$. On the other
hand, when $I_i<< w$, $\alpha$ is small and only states near the Fermi momentum 
are redistributed.

Eq. (26) shows that when $\psi_i$'s are the Bloch wavefunctions 
$\psi_{\mu \k}$, then the $I_i$'s and hence the $\Delta_i$'s will be energy
and $\k$-dependent, and the resulting correlation correction to the EMD 
will in general be {\em anisotropic}. Concerning the break at the 
Fermi momentum, solutions for some purely repulsive model interactions 
indicate that $\Delta_{\mu}(\k)$ can oscillate as a function of 
energy, and become antisymmetric with respect $\varepsilon_i$, 
going to zero at the Fermi level.\cite{hpa} If so, 
the present scheme admits the presence of discontinuities 
at the Fermi momentum in the EMD. Notably, the 
AGP satisfies the translational-symmetry requirement of 
Eq. (\ref{eq:trans}).

\section{
Illustrative 
Results on
L\lowercase{i}, B\lowercase{e} and A\lowercase{l}}
In this section, we briefly discuss some of the salient features 
of the EMD predicted by the relatively simple theoretical model of the 
preceding subsection. Examples of Li, Be and Al are considered 
as a means of getting a handle on correlation effects for a range 
of electron concentrations-- varying from one electron 
per atom in Li to two in Be and three in Al. As emphasized already in 
the introduction, our main purpose here is to 
establish that the present scheme is 
capable of explaining, at least in principle, some of the 
key experimental observations in Li, Be and Al
\cite{li1,li2,li3,li4,be1,be2,be3,al1,al2,alli}. 

In connection with the Compton spectra, it should 
be noted that the Compton profile (CP), $J(p_z)$, represents 
the double integral of the ground-state EMD
$\rho ({\it\mbox{\boldmath$p$}})$:
\begin{equation}
\label{eq_cp}
J(p_z) = \int\!\!\!\int\rho ({\it\mbox{\boldmath$p$}})dp_x dp_y,
\end{equation}
where $p_z$ lies along the scattering vector of the x-rays.

Li is considered first. Using Eq. (\ref{eq:janak}), we obtain
$I_i\approx 0.1$ Ry, 
which when compared with the valence electron band width, 
$w\approx 0.25$ Ry, yields 
$\alpha \sim I_i/w=0.4$. 
This implies that the IPM values of occupation numbers will be 
modified substantially by correlations. Fig. 2 compares a typical
CP (results along [100] are given) in Li for LDA with and without 
the LP correction, and the AGP model computation. The LDA (solid) 
displays a cusp around 0.6 a.u. which reflects 
the break in the EMD at $p_F$. The addition of 
the LP correction (light dashed) shifts some spectral weight 
from low to high momenta, but the cusp at $p_F$ is essentially 
unchanged. In sharp contrast, in the AGP curve (thick dashed), the 
feature at $p_F$ is completely smoothed out and the 
EMD is redistributed throughout the momentum space. The AGP 
results are at least qualitatively similar to the behavior of
the experimental CP's in Li which show that the size of the 
discontinuity $Z_k$ at $p_F$ is very small and that the measured
CP is substantially lower than LDA around ${\bf p}=0$
\cite{li1,li2,li3,li4}. 
We should keep 
in mind nevertheless the fact that the AGP results in Fig. 2 invoke 
a series of approximations which might oversmooth the discontinuity
at $p_F$. Notably, spin-polarized electronic 
structure computations indicate the importance of spin-wave 
fluctuations in Li\cite{tjli}, suggesting that some renormalization 
of the occupation numbers in Li could also be produced by such a 
mechanism. 

Interestingly, an extensive study of $\rm Li_{100-x}Mg_x$ 
disordered alloys \cite{li3} over the composition range $0\le x\le 40$ 
indicates that the measured CP's come into better accord 
with the LDA predictions with increasing Mg concentration. 
This observation will find a natural explanation in our 
theoretical framework since we expect the width of the 
valence band to increase with increasing Mg content with little 
change in the exchange integral $I_i$, so that the parameter $\alpha$ 
(see, Eq. \ref{eq:alpha}) will presumably 
decrease as will the correlation effect. 
%
%\begin{figure}[htb]
%\unitlength=1cm
%\begin{center}
%\begin{picture}(7,8.3)
%\put(-2.0,-1.0){\epsfysize=9cm
%\epsffile{./CP_LI.eps}}
%\end{picture}
%\end{center}
%\vskip 0.5cm
%\caption{
%Theoretical results for the [100] Compton profile in Li are 
%compared for three different models: LDA without
%LP correction (solid), LDA with LP correction (light dashed), and 
%the AGP scheme discussed in the text (heavy dashed). 
%}
%\label{CP_LI}
%\end{figure}
%

Turning to Be, the exchange integral is computed to be, $I_i\approx 0.1$ Ry, 
but $w\approx 0.8$ Ry, giving $\alpha\approx 0.13$. 
Therefore the correlation effect on the 
EMD in Be is smaller than Li, consistent with experimental 
observations \cite{be1,be2,be3}.
Fig. 3 focuses on the difference, 
$\Delta J^{Theory}=J_{AGP}-J_{LDA}$, between the AGP and the 
LDA (including the LP correction) predictions of CP's along two different 
crystal directions. The LP correction (dashed) is seen to mimick 
the overall shape of $\Delta J^{Theory}$. However, in contrast to the 
LP correction which is {\em isotropic}, the residual correlation 
correction given by $\Delta J^{Theory}$ is {\em anisotropic}. Along [100], 
$ \Delta J^{Theory}$ differs relatively little from the LP curve, but along 
[001] the excursions are much larger. The results of Fig. 3 are in 
qualitative accord with some of the characteristics of the experimental 
CP's in Be. By comparing the residuals of our Fig. 3 
with the experimental residuals (defined as,  
$\Delta J^{Expt}=J_{Expt}-J_{LDA}$, which are 
given in Fig. 4 of Ref. \cite{be3}), it will be seen that 
$\Delta J^{Expt}$, 
like $\Delta J^{Theory}$, deviates little from the LP curve along 
[100]; in sharp contrast, along [001] $\Delta J^{Expt}$ contains 
a remarkable pattern of undulations rather similar to that seen
in $\Delta J^{Theory}$ in Fig. 3. Further study is needed to pin down 
the origin of specific structures in $\Delta J^{Theory}$ in terms, for 
example, of the Fermi surface features. 
%
%\begin{figure}[htb]
%\unitlength=1cm
%\begin{center}
%\begin{picture}(7,8.3)
%\put(-2.0,-1.0){\epsfysize=9cm
%\epsffile{./DCP_BE.eps}}
%\end{picture}
%\end{center}
%\vskip 0.5cm
%\caption{
%Effect of correlation on the Compton profile 
%along two different directions in Be. The solid curve gives the
%residual difference $\Delta J^{Theory}=$ ?? between the LDA and 
%the AGP profiles; note, that $J_LDA$ here includes the LP correction, 
%which is also shown separately by dashed lines for reference. 
%}
%\label{DCP_BE}
%\end{figure}

Finally, in the case of Al, we find $I_i\approx 0.05$ Ry, which as 
expected is small, reflecting the decrease
of $I_i$ with increasing electron density.
On the other hand, the band width $w\approx 1$ Ry is quite large, so
that $\alpha\approx 0.05$ is rather small. Therefore, the EMD in Al will
not suffer much redistribution beyond the LDA via electron correlations. 
This again is consistent with experimental observations which show that
the EMD in Al is described reasonably well by the conventional LDA 
picture \cite{al1,al2,alli}. We illustrate this point further by Fig. 3 where the LDA and 
AGP momentum densities $\rho(p)$ are compared along a typical 
direction (note these are 3D momentum densities and not the CP's). 
The AGP and LDA are seen to be quite close overall, some rounding 
of the Fermi break in AGP notwithstanding. 

%
%
%\begin{figure}[htb]
%\unitlength=1cm
%\begin{center}
%\begin{picture}(7,8.3)
%\put(-2.0,-1.0){\epsfysize=9cm
%\epsffile{./MD_AL_110.eps}}
%\end{picture}
%\end{center}
%\vskip 0.5cm
%\caption{
%3D momentum density along [110] is compared in Al for the LDA and the 
%AGP case. 
%}
%\label{MD_AL}
%\end{figure}

\section{Summary and Conclusions}
\label{sec:conc}

We discuss the theoretical treatment of the 
EMD in the correlated electron gas with the purpose
of developing an understanding of recent high resolution Compton 
scattering experiments on a variety of materials. EMD can 
be expressed rigorously in terms of the eigenfunctions and eigenvalues 
of the one-particle density matrix, which play the roles 
of effective orbitals $\psi_i$'s -- the `natural orbitals', 
and the associated occupation numbers $n_i$'s, respectively. 
Much of the existing EMD work in crystals is based on 
the use of Bloch wavefunctions computed 
within the LDA and the 
related IPM values of occupation 
numbers (i.e., 1 for filled and 0 for empty states). 
In the DFT, the expectation value of any 
operator (excepting the electron density $n(\bf{r})$) must be 
corrected with respect to its IPM value in order to include 
the effect of correlations in the interacting system. The 
standard calculation of such an LP correction to 
the IPM momentum density involves the momentum 
density $\rho_h(p)$ in the interacting HEG. 
Although many attempts to evaluate $\rho_h$ in the 
HEG have been made, the recent QMC results 
are perhaps the most accurate. In this connection, we provide a 
convenient parametrization of the QMC data for $\rho_h$ as a 
function of the electron density parameter $r_s$; this form of
$\rho_h$ should be used for computing the LP correction. Note 
that the LDA-based LP correction possesses fundamental limitations, 
e.g., it can only describe an isotropic redistribution of the EMD. 

In order to go beyond the framework of the DFT, we propose using a
BCS-like wavefunction in which the starting point is an electron 
singlet pair or a 'geminal', and the many body 
wavefunction is constructed as an AGP. 
Although such an approach has been invoked previously in a variety 
of problems, we are not aware of its application to treat the 
EMD in extended systems. The AGP wavefunction tends to minimize 
the effects of the exclusion principle and possesses a fundamentally 
different nodal structure compared to the Slater-Jastrow type many 
body wavefunctions based on single-particle orbitals which are 
implicit in the LDA and QMC work. We begin to explore the nature of 
the EMD in the AGP scheme in this article by approximating the 
natural orbitals by the Kohn-Sham orbitals. By making the assumption 
that the energy gained through the BCS-like term will be roughly 
compensated by the exchange-correlation part of the Hartree-Fock term 
in the total energy functional, we show that the BCS energy scale, 
$\Delta_i\sim I_i$, where $I_i$ is a readily computed exchange-type 
integral. The key parameter which controls the redistribution of 
states in our theory then is, $\alpha\sim I_i/w$, with $w$ being the
valence electron band-width. 

Finally, we consider the application of the AGP method to discuss 
the EMD and Compton spectra of Li, Be and Al as illustrative examples. 
In Li, we find $\alpha\sim 0.4$, implying a substantial renormalization 
of the LDA states, in essential accord with experimental results which 
indicate the break in the EMD to be renormalized to a nearly zero value. 
In Be, $\alpha\sim 0.13$, and the theory predicts {\em anisotropic} 
correlation effects in the EMD which show a remarkable resemblance to 
the anisotropy of measured Compton profiles. In Al, we estimate, 
$\alpha\sim 0.05$, so that the correlations yield relatively little
modification of the EMD, and here again, this is consistent with the 
experimental observations which indicate that Al is described reasonably
within the conventional LDA picture. Taken together, 
these results--involving a range of electron concentrations, show the
potential of the AGP method in providing a theoretical method for 
describing correlation effects on the EMD in wide classes of materials, 
although further work is necessary in this connection. 
Other open questions concern the excitation properties of the AGP ground state.

\acknowledgements
This work is supported by the US Department of Energy under contract
W-31-109-ENG-38 and benefited from the allocation of supercomputer 
time at the NERSC and the Northeastern University Advanced 
Scientific Computation Center (NU-ASCC).

\newpage
\centerline{\bf Figure Captions}
\vskip 1 cm
{\noindent FIG. 1. Momentum density $\rho(p)$ in the homogeneous electron gas. 
The momentum $p$ is given in scaled units of the free electron Fermi
momentum $p_F$. Parameterized QMC (solid) and LDA (dashed) results 
are shown for $r_s$
values of 2 and 5, as discussed in the text. 
}
\vskip 0.3 cm

{\noindent FIG. 2.
Theoretical results for the [100] Compton profile in Li are 
compared for three different models: LDA without
LP correction (solid), LDA with LP correction (light dashed), and 
the AGP scheme discussed in the text (heavy dashed).}
\vskip 0.3 cm

{\noindent FIG. 3.
Effect of correlation on the Compton profile 
along two different directions in Be. The solid curve gives the
residual difference, $\Delta J^{Theory}=J_{AGP}-J_{LDA}$, between the LDA and 
the AGP profiles; note, that $J_{LDA}$ here includes the Lam-Platzman
correction, which is also shown separately (dashed lines) for reference. }
\vskip 0.3 cm

{\noindent FIG. 4.
3D momentum density along [110] is compared in Al for the LDA and the 
AGP case. 
}
\vskip 10 cm
%%%%%%%%%%%%%% Figures %%%%%%%%%%%%%%%%%%%%%%%%
\newpage
\begin{figure}[htb]
\unitlength=1cm
\begin{center}
\begin{picture}(7,8.3)
\put(-2.0,-1.0){\epsfysize=9cm
\epsffile{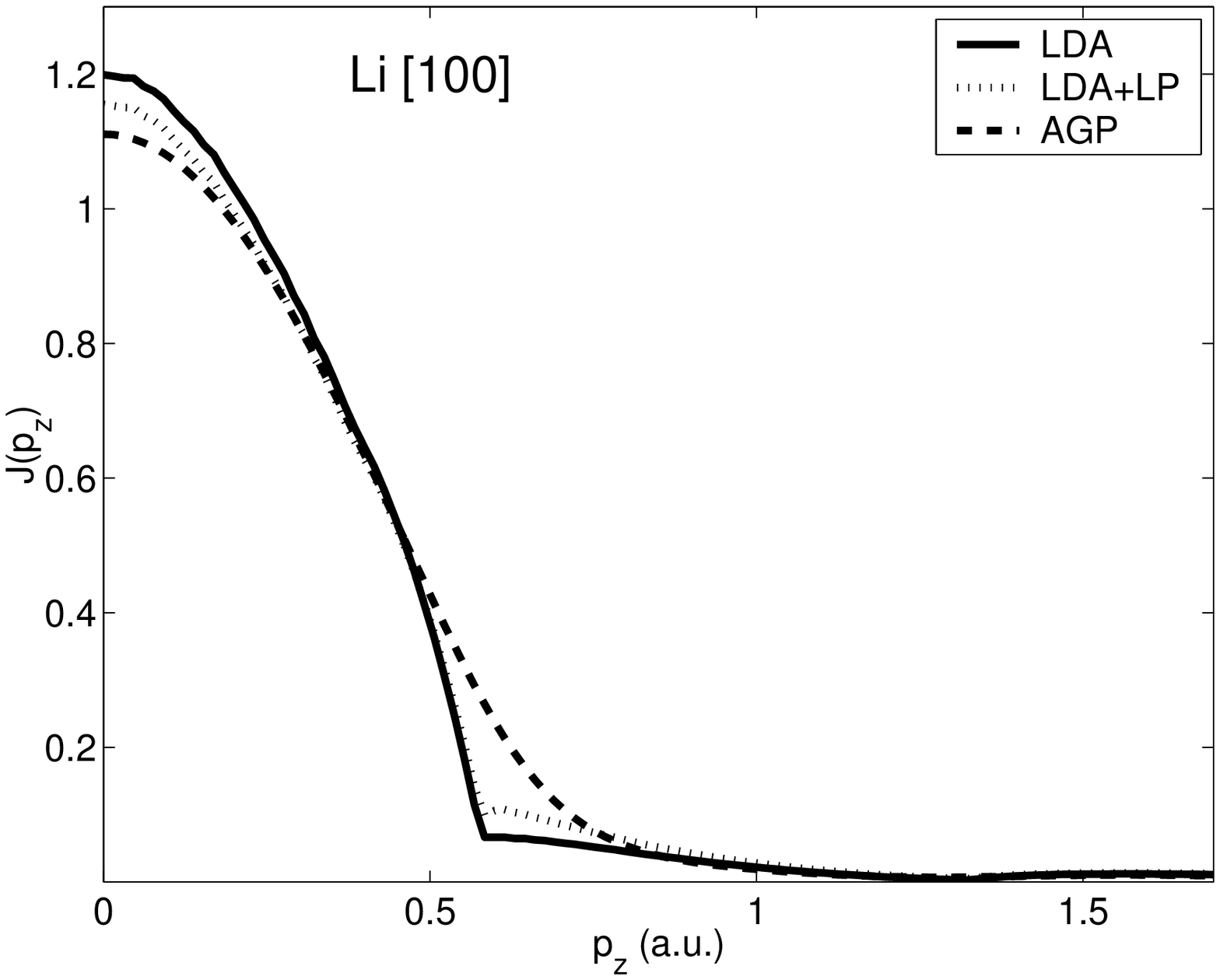}}
\end{picture}
\end{center}
\vskip 0.5cm
\caption{\ }
%\caption{
%Theoretical results for the [100] Compton profile in Li are 
%compared for three different models: LDA without
%LP correction (solid), LDA with LP correction (light dashed), and 
%the AGP scheme discussed in the text (heavy dashed). 
%}
\label{CP_LI}
\end{figure}
\newpage
\begin{figure}[htb]
\unitlength=1cm
\begin{center}
\begin{picture}(7,8.3)
\put(-2.0,-1.0){\epsfysize=9cm
\epsffile{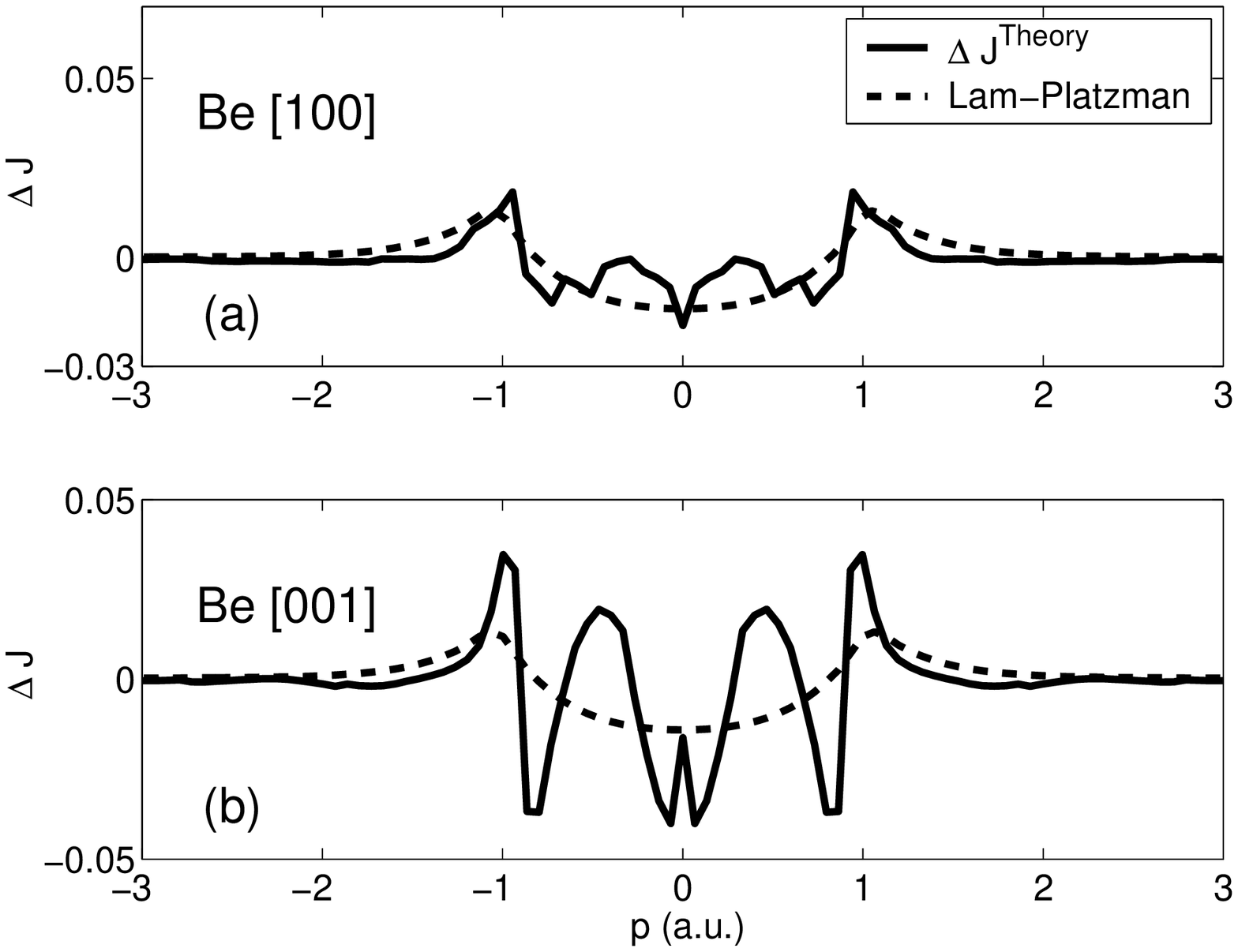}}
\end{picture}
\end{center}
\vskip 0.5cm
\caption{\ }
%\caption{
%Effect of correlation on the Compton profile 
%along two different directions in Be. The solid curve gives the
%residual difference $\Delta J^{Theory}=J_{AGP}-J_{LDA}$ between the LDA and 
%the AGP profiles; note, that $J_{LDA}$ here includes the Lam-Platzman
%correction, which is also shown separately (dashed lines) for reference. 
%}
\label{DCP_BE}
\end{figure}
\newpage
\begin{figure}[htb]
\unitlength=1cm
\begin{center}
\begin{picture}(7,8.3)
\put(-2.0,-1.0){\epsfysize=9cm
\epsffile{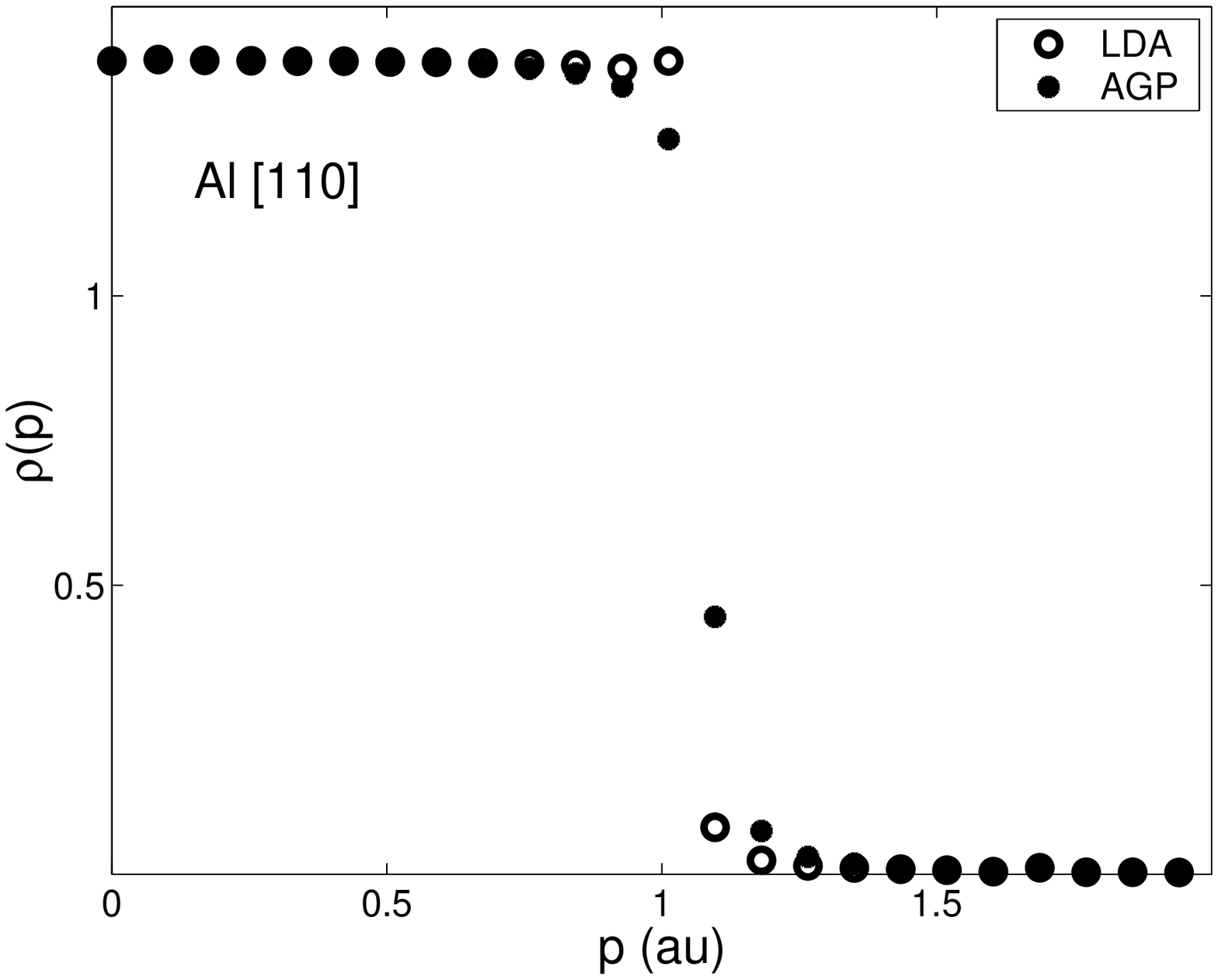}}
\end{picture}
\end{center}
\vskip 0.5cm
\caption{\ }
%\caption{
%3D momentum density along [110] is compared in Al for the LDA and the 
%AGP case. 
%}
\label{MD_AL}
\end{figure}
\end{document}